\documentclass[submission, Phys]{SciPost}
\hypersetup{
    colorlinks,
    linkcolor={red!50!black},
    citecolor={blue!50!black},
    urlcolor={blue!80!black}
}

\usepackage{epsfig,amsmath,amssymb,color,dsfont,upgreek,physics}
\usepackage{mathrsfs}
\usepackage{mathtools}
\usepackage{bbold}
\usepackage{adjustbox}
\usepackage{comment}
\usepackage{float}
\usepackage{braket}
\usepackage{listings} % To Show Pseudo and Actual Code
\lstdefinestyle{mystyle}{
    backgroundcolor=\color{backcolour},
    commentstyle=\color{codegreen},
    keywordstyle=\color{magenta},
    numberstyle=\tiny\color{codegray},
    stringstyle=\color{codepurple},
    basicstyle=\ttfamily\footnotesize,
    breakatwhitespace=false,
    breaklines=true,
    captionpos=b,
    keepspaces=false,
    numbers=none,
    numbersep=3pt,
    showspaces=false,
    showstringspaces=false,
    showtabs=false,
    tabsize=1
}

\usepackage{xcolor} % Extra Colours
\definecolor{codegreen}{rgb}{0,0.6,0}
\definecolor{codegray}{rgb}{0.5,0.5,0.5}
\definecolor{codepurple}{rgb}{0.58,0,0.82}
\definecolor{backcolour}{rgb}{0.95,0.95,0.92}

\newcommand{\be}{\begin{equation}}
\newcommand{\ee}{\end{equation}}
\newcommand{\lra}{\leftrightarrow}

\DeclareSymbolFont{usualmathcal}{OMS}{cmsy}{m}{n}
\DeclareSymbolFontAlphabet{\mathcal}{usualmathcal}

\begin{document}

\begin{center}{\Large \textbf{
Quantum Monte Carlo simulations in the restricted Hilbert space of
Rydberg atom arrays
}}\end{center}

% TODO: write the author list here. Use first name (+ other initials) + surname format.
% Separate subsequent authors by a comma, omit comma and use "and" for the last author.
% Mark the corresponding author with a superscript star.
\begin{center}
Pranay Patil\textsuperscript{1$\star$}
\end{center}

\begin{center}
{\bf 1} Max-Planck Institute for the Physics of Complex Systems,
Dresden, Germany
\\
${}^\star$ {\small \sf pranay.patil15@gmail.com}
\end{center}

\begin{center}
\today
\end{center}

% For convenience during refereeing (optional),
% you can turn on line numbers by uncommenting the next line:
%\linenumbers

\section*{Abstract}
{\bf
% TODO: write your abstract here.
Rydberg atom arrays have emerged as a powerful platform to simulate a number
of exotic quantum ground states and phase transitions. To verify these
capabilities
numerically, we develop a versatile quantum Monte Carlo sampling technique
which operates in the reduced Hilbert space generated by enforcing the
constraint of a Rydberg blockade. We use the framework of stochastic series
expansion and show that in the restricted space, the configuration space
of operator strings can be understood as a hard rod gas in $d+1$ dimensions.
We use this mapping to develop cluster algorithms which can be visualized as
various non-local movements of rods. We study the efficiency of each of our
updates individually and collectively. To elucidate the utility of the
algorithm, we show that it can
efficiently generate the phase diagram of a Rydberg atom array, to temperatures
much smaller than all energy scales involved, on a
Kagom\'e link lattice. This is of broad interest as the presence of a $Z_2$
spin liquid has been hypothesized recently \cite{Vryd}.
}

\vspace{10pt}
\noindent\rule{\textwidth}{1pt}
\tableofcontents\thispagestyle{fancy}
\noindent\rule{\textwidth}{1pt}
\vspace{10pt}

\section{Introduction}
\label{sec:intro}
% TODO: write your article here.
Recent advances in ultra-cold atom arrays in optical traps have generated
renewed interest in understanding the possible condensed matter systems
which can be realized in such systems. One of the most prominent of these
is the Rydberg blockade mechanism, which is engineered by exciting atoms
with a single occupancy in the outer $s$-shell to high principal quantum
numbers~\cite{jaksch2000fast,lukin2001dipole,urban2009observation,browaeys2020many}.
The large electron cloud of such a state generates a strong
Coulomb repulsion, which prevents the excitation of atoms in close
vicinity. This can be modeled theoretically as an exclusion process,
and has led to the so-called $PXP$ model.
This strongly interacting model
of spins has a variety of interesting features in both its
static and dynamic behavior, and has been found to be of interest for
frustrated magnetism, quantum many-body scar states and quantum
computing
~\cite{turner2018weak,shiraishi2019connection,iadecola2020quantum,surace2020lattice,wurtz2023aquila}.

Within the context of magnetism, it has been found that the most interesting
consequences of this exclusion principle are seen in two- or three-dimensional
systems. For the strongly correlated regimes of interest for realizing
non-trivial magnetic order, exact analytic treatments are not feasible, and
one of the most powerful numerical techniques to simulate such systems is
quantum Monte Carlo (QMC). Although the models for Rydberg atoms are similar to
transverse field Ising (TFI) models,
and thus lack the infamous sign problem~\cite{henelius2000sign} which
makes QMC unscalable, the phase space structure allowed by the Rydberg
blockade is highly correlated and notoriously hard to sample efficiently.
There have been systematic developments in the past
~\cite{sandvik2003stochastic,melko2013stochastic}
to sample TFI (including the highly frustrated case
~\cite{kandel1990cluster,zhang1994cluster,wu2021z,patil2019obstacles})
and related models such as the quantum dimer model within the framework
of stochastic series expansion (SSE-QMC).
Of particular relevance to Rydberg atom arrays emulating frustrated magnets
are the motif-marking methods\cite{biswas2016quantum},
the sweeping cluster algorithm\cite{yan2019sweeping,yan2022triangular},
local in space cluster updates\cite{biswas2016quantum,patil2019obstacles},
and directed loop updates\cite{syljuaasen2002quantum,syljuaasen2003directed}.
Although these algorithms have been implemented for Rydberg atom arrays
\cite{yan2023emergent,yan2022triangular,Melko},
the parameter ranges which have been investigated have been relatively
modest due to inefficiencies in sampling, which are especially acute
at the low temperatures required to see genuine quantum many-body physics.

Here we develop a formalism for SSE-QMC directly in the restricted space
generated by the Rydberg blockade by realizing a mapping between the operator
string sampled by SSE-QMC and a hard rod model. Versions of the latter
have already been studied from the perspective of soft matter physics, where
they have been often used to model nematic crystals
~\cite{kundu2016phase,gschwind2017isotropic}.
Due to the mapping from
a quantum model, the hard rod model generated in our case is fundamentally
anisotropic, and requires new updates which we describe here.
Updates in a similar rod language were also used for a transverse
field Ising glass in the context of continuous time QMC
\cite{rieger1999application}.

We begin by describing the Hamiltonian and operators relevant for SSE-QMC,
and present the $d+1$-dimensional hard rod model realized in space-time.
This is followed by a description of the standard diagonal update, and
a detailed discussion of three cluster updates in the rod language which allow
us to move efficiently within the phase space of allowed rod configurations.
To enhance the efficiency of the cluster updates described, we implement
an optimized form of parallel tempering, and provide detailed results on
the efficiency of various updates in different regions of parameter space.
We follow this up
by showing an application of our algorithm to the case of a Kagom\'e link
lattice, which has recently been proposed to realize a $Z_2$ quantum spin
liquid, and study the dependence of relevant order parameters on temperature and
the detuning parameter. Our preliminary results find no presence of spin
liquid behavior
down to temperatures which are $5\%$ of the smallest energy scale in the
system. We conclude by emphasizing that our algorithm is lattice independent
and by elucidating other cases of interest where it can be applied.

\section{Model and formulation of SSEQMC}

The most common model\cite{Vryd,yan2023emergent}
for Rydberg atom arrays is given by hard-core bosons
with a strong repulsive interaction and a kinetic term which
creates/annihilates bosons. We define our system on a 2D lattice with $N$ sites,
and do not specify a particular geometry for our analysis
(this can be seen as a 2D version of the model studied in Ref.~\cite{fendley2004competing}).
For ease of comparison we parameterize
the Hamiltonian in the same way as Ref.~\cite{Vryd}, and reproduce it below
for completeness :
\be\label{Hryd}
H=\frac{\Omega}{2}\sum_i(b_i+b_i^{\dagger})-\delta\sum_i n_i+
\frac{1}{2}\sum_{i,j} V(|i-j|)n_in_j.
\ee
The Coulomb repulsion due to the Rydberg blockade is encoded in $V(|i-j|)$,
which is known to have the functional form $V(r)\approx 1/r^6$ \cite{urban2009observation}.
As this is extremely short range, we make the approximation that for
nearest neighbors $V(r)\to\infty$, and $V(r)=0$ otherwise. This provides
us with an effective model to understand the effect of the Rydberg blockade.
We have effectively removed all states from our boson occupation basis
which have bosons on nearest neighbors. Within this restricted space, the
Hamiltonian simply reduces to
$H=\frac{\Omega}{2}\sum_i(b_i+b_i^{\dagger})-\delta\sum_i n_i$.
It is important
to note here that even though it is not apparent from the Hamiltonian, the
restricted Hilbert space renders the system genuinely interacting.
As there is a freedom of one energy scale, we set $\delta=1$ for
all our numerical results, unless otherwise mentioned.
In the restricted space, the action of the chemical potential is to maximize
the density of bosons, thus leading to fully packed configurations with
no neighboring sites simultaneously occupied by bosons. In contrast, the
kinetic term acts by creating resonances between the occupied and
unoccupied state, and in many cases leads to a phase continuously connected
to the trivial $x$-polarized paramagnet one would expect for a
TFI model in the limit of large transverse field.

\begin{figure}[t]
\includegraphics[width=\hsize]{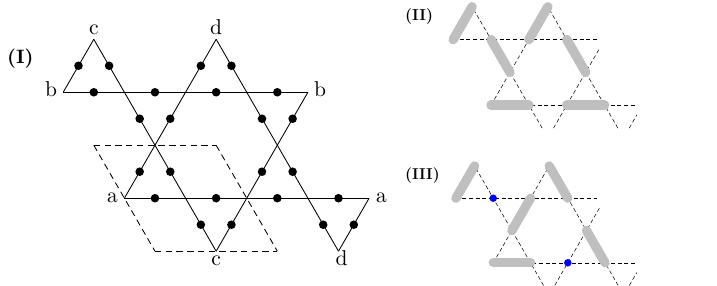}
\caption{
(I) Kagom\'e link lattice with sites (highlighted) living on the links.
A single unit cell has 6 sites and is shown with a dashed line border.
Periodic boundary conditions are indicated with pairs of alphabets.
(II) Sites occupied by bosons represented by a fully packed hard-core
dimer configuration which respects the Rydberg constraint.
(III) Partially filled dimer configuration with some sites on the Kagom\'e
lattice (shown with blue filled circles) not covered by dimers.
}
\label{FigKL}
\end{figure}

Stochastic series expansion~\cite{sandvik2003stochastic,sandvik1992generalization}
proceeds by expanding the exponential in the
partition function $Z=Tr[e^{-\beta H}]$ as
$\sum_n\frac{(-\beta)^n}{n!}\sum_{s_n}a_{s_n}Tr[s_n]$, where $s_n$ lists all
length-$n$ operator strings of the form $Tr[H_{s^1_n}H_{s^2_n}...H_{s^n_n}]$,
and $a_{s_n}=(\Omega/2)^{n_K}(-\delta)^{n_V}$ denotes the weight generated by
the coefficients in the Hamiltonian. Here, $n_K$ and $n_V$ denote the number of
kinetic $(b_i+b_i^{\dagger})$ and potential $(n_i)$ operators in the string
($n_K+n_V=n$),
and the trace is calculated by sampling over the basis states allowed by
the Rydberg blockade condition. All terms in this expansion are positive,
as the number of kinetic terms must be even to return the state to itself
(enforced by the trace), implying that $n_K$ is even and $(-\beta)^{n_K}$
is positive, and each potential term carries a negative sign,
which cancels the effect of the negative sign coming from $(-\beta)^{n_V}$.
We use the boson occupation basis for computations, and in this basis the
actions of these two types of operators can be pictorially represented as
\begin{equation}\label{Evts}
H_K=\adjustbox{valign=c}{\includegraphics[height=0.7cm]{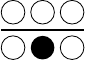}} +h.c,
\ \
H_V=\adjustbox{valign=c}{\includegraphics[height=0.7cm]{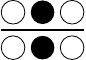}}.
\end{equation}
Note that this representation is in 1D, and for a higher dimensional
system, the operators will contain all nearest neighbors, thus
enforcing the Rydberg blockade. For the Kagom\'e link lattice shown
in Fig.~\ref{FigKL}, each site has six nearest neighbors, and thus
the operators are defined on 7 spatial sites.
In addition to these operators, we also add an energy shift to our
Hamiltonian in the form of $-(\Omega/2)I$, as we implement
updates which separately exchange $H_V\leftrightarrow I$
and $H_K\leftrightarrow I$ to ensure ergodicity. The global
shift does not change the physics and is added in purely to facilitate
convenient updates. The negative sign in the coefficient is chosen to
remove the sign problem, as already discussed for $H_V$.
For ease of computational programming, we also work with a fixed
length array to store the operator string, and ensure that we have chosen
a length $M$
which is large enough to accommodate relevant fluctuations of the number
of operators $n$. The $n$ operators are distributed randomly in $M$
``slices'' and the rest of the $M-n$ slices are left empty.
This is a standard practice for SSE\cite{sandvik2010computational},
and is essential for implementing the diagonal update.

It is instructive to also consider the construction of an operator
string by evaluating the trace using insertions of complete basis states.
This can be understood for a sample operator string $s$ of length three,
given by $Tr[H_{s^1}H_{s^2}H_{s^3}]$. Insertions of our basis projector,
$\sum_{\alpha}\ket{\alpha}\bra{\alpha}$
between every pair of operators reduces the trace to
\begin{equation*}
\sum_{\alpha_1}\sum_{\alpha_2}\sum_{\alpha_3}
\braket{\alpha_1|H_{s^1}|\alpha_2}
\braket{\alpha_2|H_{s^2}|\alpha_3}
\braket{\alpha_3|H_{s^3}|\alpha_1}.
\end{equation*}
In most formulations of stochastic series expansion, one uses a projector
onto the complete Hilbert space, i.e. the identity operator
$I=\sum_i\ket{i}\bra{i}$, whereas we restrict the basis states
allowed. This restriction implies that not all non-zero matrix elements
of the identity $I$, which we have added to the Hamiltonian,
are accessed.

\begin{figure}[t]
\includegraphics[width=\hsize]{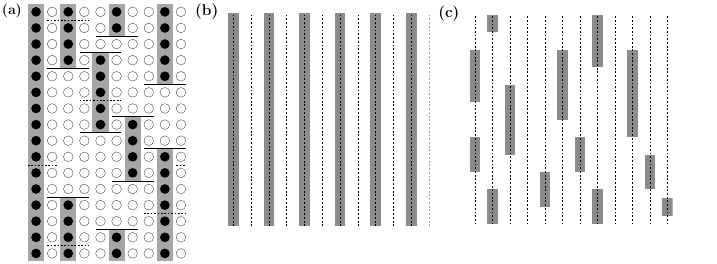}
\caption{Space-time configurations for a periodic 1D chain with
the vertical direction corresponding to imaginary time being also periodic :
(a) Operator string using the operator representation defined in Eq.~\eqref{Evts},
and corresponding rod configuration in gray;
(b) Rod configurations dominating in the $\Omega=0$ limit are made up of
fully packed time-spanning rod configurations with no ends;
(c) In the $\Omega\gg\delta$ limit, the rods are short and uncorrelated.
}
\label{FigB}
\end{figure}

Using this notation each operator string can be represented as a boson
occupation diagram in one higher dimension. An example of this is shown in
Fig.~\ref{FigB} for a one-dimensional system. Note that due to the Rydberg
constraint which prevents simultaneous occupation of neighboring sites,
regions of the operator string diagram which correspond to occupied sites
can be identified as rods (shown with rectangles with dashed borders
in Fig.~\ref{FigB})
with an exclusion radius of one lattice spacing. Thus configurations of
operator strings can be viewed as arrangements of vertical ``thick'' rods.
To get an intuition of the mapping, let us consider the classical
($\delta\gg\Omega$) and quantum ($\Omega\gg\delta$) limits. In the former,
the operator string is dominated by $H_V$ and the rods tend to be long
as the presence of end points ($H_K$) is suppressed. In addition, $H_V$
promotes a maximum coverage of space-time by rods, leading to maximally
packed configurations. The length of the
operator string is controlled by the inverse temperature $\beta$ and in
the classical limit, the rods span the entire system in the vertical direction.
In contrast, a high density of $H_K$ promotes short rods, and for many
lattice geometries, it may be possible to realize a rod paramagnet, where the
density of rods is low and their positions are uncorrelated. These limits
are shown pictorially in Fig.~\ref{FigB}. The sampling of operator strings can
now be considered to be a sampling over rod configurations, and we describe
in the following section various updates which we use to carry out this
sampling, along with benchmarks of efficiency for the novel updates we
have introduced. To understand the inherent non-trivial nature of these updates,
one can consider a highly dense network of rods which are randomly placed, but
respect the Rydberg constraint, and attempt to imagine a controlled modification
of rods which will lead to another valid rod configuration. A quick thought
experiment is sufficient to convince ourselves that this is not a trivial
exercise, especially given that a large penalty in probability must be paid,
due to the chemical potential $\delta$,
if the rod coverage is modified substantially.

\section{Monte Carlo updates and efficiency benchmarks}

Our simulations are initialized using an empty operator string, and with
all the states in the set $\{\alpha_1,...,\alpha_M\}$ set to the empty
(zero boson) state. $M$ is chosen to be $N$ initially, and grown dynamically
during the equilibration process to accommodate fluctuations of $n$,
until it saturates. We set $M=(4/3)n_{max}$, where $n_{max}$ is the
largest value of $n$ seen till the current simulation time. The factor
of $4/3$ is chosen to provide sufficient room for any further fluctuations
and to allow movement between operators in the diagonal update.
A configuration for our simulation is defined by a particular
operator string of length $M$ (including empty locations), and a
particular choice for the set $\{\alpha_1,...,\alpha_M\}$, and the probability
of this configuration is $\propto\frac{1}{^MC_n}\frac{(\beta^n)}{n!}\delta^{n_V}
\big(\frac{\Omega}{2}\big)^{n_K}$. The combinatorial factor here appears
as the expansion of $n$ to $M$ creates an unnecessary degeneracy in
the operator string configurations over $M$ slices, which correspond
to the same string. This must be suppressed to avoid spurious weight factors.
The number of equilibrium steps are chosen to be large enough, such that
further increments do not change our results, and the updates we use
are described below. Although our updates are applicable to general lattices,
for ease of understanding, we illustrate most of them in the context of a
1D chain, except for the rod diffusion update, where the 2D case requires
important considerations which are absent in the 1D case. However, the
measurements of efficiency of various updates are done on the 2D Kagom\'e
link lattice, also known as the ruby lattice. We make this choice to showcase
the applicability of our updates to frustrated systems which are expected to
be the most difficult to simulate using traditional algorithms.

The ruby lattice (shown in Fig.~\ref{FigKL})
is generated by considering Rydberg atoms living on the links
of a Kagom\'e lattice\cite{Vryd}.
An excited Rydberg atom can be understood as a dimer occupying the link
under consideration, and the Rydberg constraint then reduces to a hard-core
dimer constraint, which requires that dimers do not share ends. In the classical
limit ($\Omega=0$), the ground state can then be understood to be a fully-packed
dimer liquid with paramagnetic
correlations~\cite{domb2000phase,misguich2002quantum}.
This model is expected to host a $Z_2$ spin liquid at zero temperature
at intermediate $\Omega/\delta$.
We use a unit cell of six sites (as shown with dashed border in Fig.~\ref{FigKL}),
and set periodic boundary conditions for the underlying tilted square
lattice with $L_x=L_y=L$.
States which satisfy the Rydberg constraint can be represented as hard-core
dimer coverings; examples of the same are shown in Fig.~\ref{FigKL} for both
the fully filled and partially filled cases.

We begin here with a description of the diagonal update, which does not modify
the rod configuration underlying the operator string, but performs the crucial
insertions and deletions of diagonal operators. This is followed by three
cluster updates, which modify both the rod configuration and the operator
string, and they are presented in increasing order of spatio-temporal range.
We end by discussing the optimized parallel tempering method used to ensure
ergodicity at low temperatures. For all updates other than the diagonal one,
we present benchmarks for efficiency.
As we will exclusively study the Kagom\'e link lattice from this point
onwards, we will use dimer density to refer to link occupation (rather than
boson density as done above).

\subsection{Diagonal update}

A particular configuration during our Monte Carlo simulation is given by
a fixed $s_n$, and fixed set $\{\alpha_1,...,\alpha_M\}$. The diagonal
update for SSE~\cite{sandvik2010computational}
changes the configuration by only modifying the
operators in $s_n$ which are diagonal. This is a local operation which
does not require any modification of operators other than the one under
consideration, or of any basis state $\alpha$. As the modifications do not
depend on the status of other operators, we can carry out this update by
sequentially running once over the entire operator string.
The updating procedure is as follows.

We pick a particular location in our fixed length operator string. If this
location is occupied by a diagonal operator, we attempt to remove this
operator with probability
\be
\frac{M-n+1}{\beta N(\delta+\frac{\Omega}{2})}.
\ee
If this is greater than unity, then we interpret it as probability 1.
This follows in a straightforward manner from the detailed balance condition,
and has been explained in detail for the Heisenberg model in
Ref.~\cite{sandvik2010computational} and for the Rydberg model in
Ref.~\cite{Melko}. 

If the location in the operator string which we have picked is empty, we choose
to fill it with an operator with probability
\be
\frac{\beta N(\delta+\frac{\Omega}{2})}{M-n}.
\ee
If the decision to add an operator is accepted, we choose to pick an identity
operator with probability $P_{I}=(\Omega/2)/(\delta+\Omega/2)$,
or an $H_V$ operator with probability $1-P_{I}$. Once an operator
is chosen it is assigned to a random site on the lattice. If we chose an
$H_V$ operator, then we check if the local background at the chosen site is
consistent with the pattern required by $H_V$ (as shown in Eq.~\eqref{Evts}),
and if so, only then we add this operator. If an identity operator is chosen,
there is no such condition, and we always add it. The procedure discussed
here closely follows the diagonal updates introduced for the SSE framework
for the Heisenberg model.

\subsection{Local segment update}\label{SLS}

\begin{figure}[t]
\includegraphics[width=\hsize]{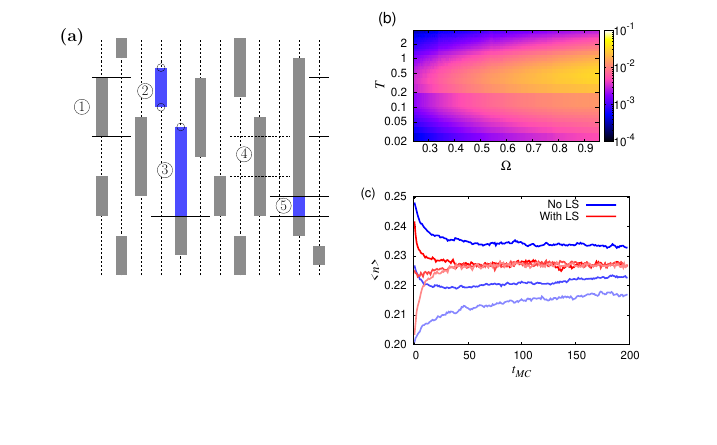}
\caption{
(a) Illustration of the possible moves which can be engineered within the
local segment update : 1. A rod segment bounded by two $H_K$ operators
can be removed and the operators replaced by $I$ operators (represented
as circles); 2. The inverse of process 1, the empty region targeted is
highlighted in blue; 3. An empty region, between an $H_K$ and $I$
operator, which cannot be flipped as it would conflict with a nearest
neighbor rod; 4. Rod segment between two $H_V$ operators can be flipped
to an empty space bounded by two $H_K$ operators; 5. Inverse of process 4.
(b) Average success probability of the local segment update.
(c) Decay of dimer density ($\braket{n}$) as a function of Monte Carlo
time for three different initial conditions with and without the local
segment update ($L=4,\Omega=0.6,T=0.15$).
Error bars are of the order of the temporal fluctuations
and are suppressed for clarity.
}
\label{FigLS}
\end{figure}

This update can be understood in the rod language as local deletions/insertions
of rod segments. These segments are generically expected to be short in the
imaginary time direction for most of parameter space, and thus this update
is local in space (acting only on a single site) and local in time (modifies
the state of the corresponding site on only a few slices). This is implemented
in two ways, as described in App.~\ref{AppA},
and shown schematically using representative
examples in Fig.~\ref{FigLS}(a).

Without the Rydberg constraint, this update would be identical to the single
line updating step for the transverse field Ising
model~\cite{rieger1999application}. However, as our model is defined only
within the Hilbert space where the Rydberg constraint is not violated,
introducing a new rod segment can only be done if it is not adjacent to
another rod at the same slice in imaginary time. This is illustrated in
Fig.~\ref{FigLS}(a), where rod position 2 is a valid insertion,
whereas position 3 is not due to an overlap with its neighbor. This check
does not need to be performed while removing a rod as the Rydberg
constraint only prohibits the existence of neighboring rods and puts
no constraint on the position of absent rods.

At high temperatures, the operator strings are short and can be intuitively
understood to be nearly absent of operators. In this limit, we still need
an update which is able to sample the entire Rydberg allowed state space.
This is done by picking a site at random and checking that the entire column
corresponding to it is operator free. If so, we flip the status of this column
after ensuring that there is no conflict created by this flip. This mechanism
trivially ensures completely ergodicity at infinite temperature.

To gain an intuition of the efficiency of this update, let us consider the
classical limit $\delta\gg\Omega$. In this limit, as we have already discussed
the rod configuration is dominated by long rods that break spatial symmetry.
In this limit, the rod coverage is well defined and incompressible, i.e. the
configurations which dominate the partition function have almost identical
rod density, which is the highest attainable without violating the Rydberg
constraint. This implies that any update which changes rod density is unlikely
to succeed due to the small probability of the configurations it creates. In
the opposite limit, $\delta\ll\Omega$, the rod configurations have large
numbers of rod ends and no preference for rod coverage, leading to high
probabilities for our local segment update. This can also be directly
understood by studying the average compressibility,
$\kappa=\braket{n_i^2}-\braket{n_i}^2$, as a function of $\Omega$ for fixed
$\delta$, and indeed it is known that for the Kagom\'e
link lattice~\cite{Vryd},
regions of large compressibility are present at intermediate $\delta/\Omega$
and this is where we find that our update is highly successful.

To ensure that the update samples over the entire operator string on average
in each Monte Carlo step, we track the average area of space time that the
update covers on each attempt, and choose the number of repetitions of the
update to cover all of space time. We can also track the total size of the
regions that are updated successfully. Although these regions can overlap
as each initial position is chosen randomly, this size as a fraction of
the full space-time can be taken as a proxy for the success probability.
This gives us a metric for the
efficiency of our update, and our results are shown for a range of
temperature and $\Omega$ in Fig.~\ref{FigLS}(b). Though this update is
expected to be highly successful in the $\Omega\gg\delta$ regime,
we still find a small but sufficient success rate for intermediate
$\Omega$ and $T$. In the limit of $\delta=0$ and finite $\Omega$,
a similar local update~\cite{greitemann2018scipost} has been used
to successfully study the PXP model on the square lattice
\cite{yue2021order}.
As argued from the compressibility, we find that the success rate
is much smaller in the frustrated regime
$(\delta\gg\Omega)$. This is seen especially clearly at low temperatures,
which are required to understand genuine quantum many-body effects.

As the acceptance rate is not necessarily an indicator of good sampling
for the observables of interest, we study the correlation in simulation time
of the dimer density (rod coverage discussed above) with and without
this update, in the presence of all other updates. We choose this observable
as inserting and deleting rods directly changes $n_i$, which none of the other
updates do explicitly. We choose $T=0.15$ and $\Omega=0.6$, as
Figs.~\ref{FigLS}(b),~\ref{FigVS}(b) and ~\ref{FigRD2}(a)
show that the acceptance probability for all three off-diagonal updates
which we have discussed
is close to maximum here, and study the total dimer density at a fixed
imaginary time slice ($n=\frac{1}{N}\sum_i n_i$) starting from various
initial conditions. The latter are generated using
simulations which employ all updates and are thus representative of the
equilibrium distribution. We deliberately do not use parallel tempering
here as we want to follow the loss of memory of the initial condition
using just the updates available for individual copies of the system.
In Fig.~\ref{FigLS}(c) we show that without the local update, the dimer
density does not equilibrate to a unique value as a function of Monte
Carlo time for three different initial conditions when averaged over
many realizations of the Markov chain dynamics.
This implies that the simulation is not ergodic. However, the
addition of the local update leads to the a single equilibrium value
,as seen in Fig.~\ref{FigLS}(c),
for the same initial conditions, with a relatively short Monte Carlo
time duration required to lose memory of the initial condition.
The data presented here is for $L=4$, and the same behavior is expected
to hold at higher $L$. This clearly
demonstrates the importance of this simple update.

\subsection{Vertical shuffle update}\label{SVS}

\begin{figure}[t]
\includegraphics[width=\hsize]{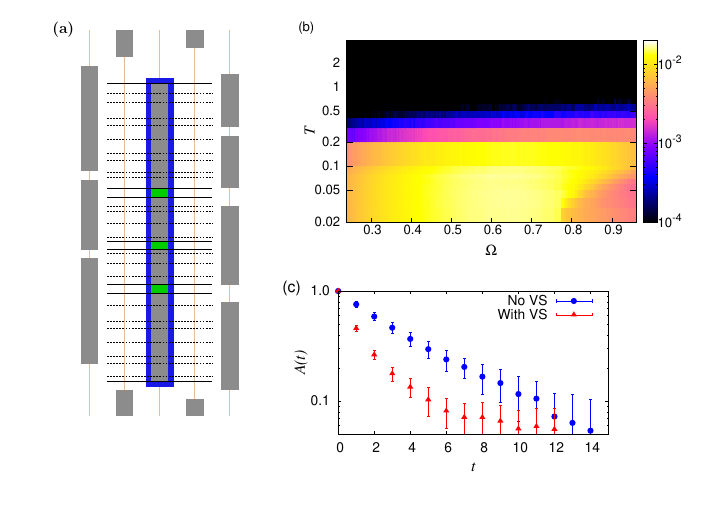}
\caption{
(a) Example of an active region, shown with a blue border, which comprises
multiple empty segments bounded by $H_K$ operators. The entire active region
for the vertical shuffle update is made up of all active regions of the
type shown here which live on the chosen spatial site.
(b) The success probability of the vertical shuffle update is seen to
be significant only at low temperatures.
(c) Autocorrelation of dimer density with and without the vertical shuffle
update for $L=4,\Omega=0.4$ and $T=0.06$.
}
\label{FigVS}
\end{figure}

As discussed above, the local segment update is inefficient in the classical
limit, where the rod coverage is close to maximal, and creation/annihilation
of segments is strongly suppressed. The rod configurations in this limit
$(\Omega\ll\delta)$ can be visualized as long rods which individually span
either the entire time direction or a large fraction of it,
thus fixing the same pattern for boson occupation for a large number of
time slices, with few uncorrelated segment breaks intervening. This is
illustrated in Fig.~\ref{FigVS}(a)
for the 1D chain, and can be easily visualized
for 2D lattices by considering any fully packed spatial arrangement.
As shown in Fig.~\ref{FigVS}(a), the number of $H_V$ type operators
far exceeds the number of $H_K$ type in the limit where this update
is expected to be efficient.
The short segment breaks can now be considered to be free objects which can
be shuffled vertically on the same spatial site within regions which do
not create rod exclusion violations. This is a micro-canonical update as
only the positions of operators is modified, not their type, leaving
the probability of the operator string unchanged. For this update, we ignore
the identity operators, and build a connected linked list of only the
$H_K$ and $H_V$ operators assuming that the identity operators are absent.
Note that during this listing process we must exclude segment breaks where
the intervening time slices host a rod on one of the spatial neighbors of
the chosen site. Such a segment is immobile as replacing it with a rod
would create two neighboring rods, which is a violation of the Rydberg
constraint.
The details of the update process are presented in App.~\ref{AppB}.

We expect vertical shuffling to be effective in the regime of low temperature
(large temporal extent) and weak quantum fluctuations. Our numerical results
for the efficiency, calculated as discussed above,
are shown in Fig.~\ref{FigVS}(b) for the Kagom\'e link lattice, and we find that
indeed in the regime discussed above, we have success rates which are $O(1)$.
At high temperatures, there are relatively few operators in space-time, thus
the required local environment is hardly ever seen by the update.
For a particular time slice, the action of this update would be to modify
the total dimer occupation one dimer at a time. The dimer density ($n_i$)
is an important physical observable and the direct impact of this update on the
estimation of $n_i$ can be understood by considering the autocorrelation
function with and without the vertical shuffle update.
We define the autocorrelation function as
\be\label{eauto}
A(t)=\sum_{\tau}\frac{1}{N}\sum_i(\braket{n_i(\tau+t)n_i(\tau)}-\braket{n}^2),
\ee
where $t,\tau$ denote simulation time, and $n=\frac{\sum_i n_i}{N}$ is the
dimer density. The autocorrelation is calculated using the
spatial configuration on a random reference slice, and the location of
this reference slice is kept fixed for all simulations.
We choose $\Omega=0.4$ and $T=0.06$, as this is a region where the
vertical shuffle update is expected to have
a high success probability (Fig.~\ref{FigVS}(b))
and the local segment update has a low to medium probability
(Fig.~\ref{FigLS}(b)). We show our results for $A(t)$, normalized by $A(0)$
for convenient visualization, in Fig.~\ref{FigVS}(c), and find that the
autocorrelation decays faster when the vertical shuffle update is
implemented. This improvement is not drastic as we do not find
any region in our phase diagram where we have long rods with many
small segment breaks. 
However, as both the local segment and vertical shuffle updates
are local in space-time, the improvement over using just the former is not
expected to scale with system size. To see this, let us consider the
probability with which the local segment update achieves the $H_K$ pair
to $H_V$ pair swap implemented here. Replacing two $H_V$ operators by
two of $H_K$ type will be accepted with a probability given by
$(\Omega/2)^2$, which is $\ll 1$ in the limit of small $\Omega$, but
is system size independent.

\subsection{Rod diffusion update}\label{SRD}

\begin{figure}[t]
\includegraphics[width=\hsize]{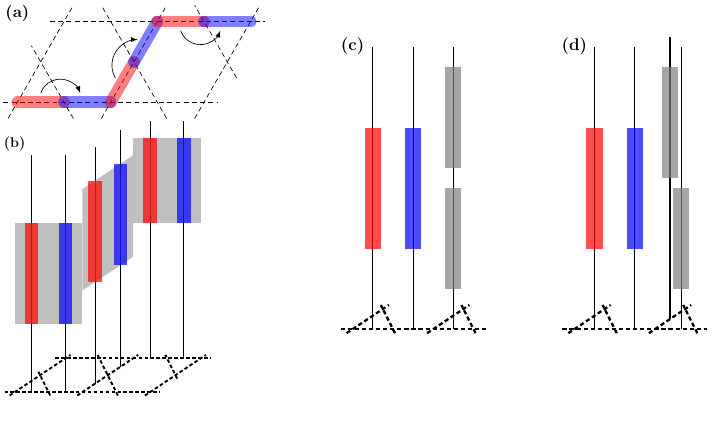}
\caption{
(a) A spatial worm update for a classical dimer model which serves as the
template for the corresponding rod diffusion update shown in (b). In all
cases, red denotes current position and blue denotes proposed position.
In (b), the membrane of modification is shown in gray, and its temporal
extent changes with spatial location (rods which do not participate in
this update are not shown to avoid clutter).
(c) and (d) represent cases where
the update must be aborted due to the conflicting number of rods to
be moved being $>1$.
}
\label{FigRD1}
\end{figure}

The two updates we have described above are limited in their action to a
single spatial site. This is highly restrictive and immediately suggests that
our simulation would not be able to sample configurations with different
spatial patterns efficiently. To remedy this, we introduce here a cluster
update which moves rods in both space and time. Similar to the vertical
shuffle update, this is a micro-canonical update, as we do not change the
numbers of $H_V$ or $H_K$ type operators, but only their positions.
Identity operators play no role in this update. The detailed balance
relations used here are non-trivial only for a spatial system which is at
least two-dimensional, and thus we describe this update on the Kagom\'e
link lattice (Fig.~\ref{FigKL}). This update can be thought of as the
worm update for classical dimer models~\cite{alet2006classical},
extended into the temporal dimension. Although this may suggest that
our update stays in a fixed temporal slice, this is not the case,
and it allows for additional flexibility as explained below.

We begin by creating a list of all rods and their positions in the current
space-time configuration.
This needs to be done only once if the update is to be repeated a set number
of times. After each successful update, we just edit the details of the rods
which have changed positions.
Of all available rods, we pick one at random to initiate the update.
We then proceed to sequentially move conflicting rods, thus building a membrane
of modified rods as we go, until we reach a
configuration which is a valid rod configuration. 
We use a simple solution to the detailed balance equations which
requires that there should be at most one conflicting rod at each step
of the membrane building process (Fig.~\ref{FigRD1}(b)).
A detailed discussion of this condition as well as of the recursive
steps required to move rods is presented in App.~\ref{AppC}.

\begin{figure}[t]
\includegraphics[width=\hsize]{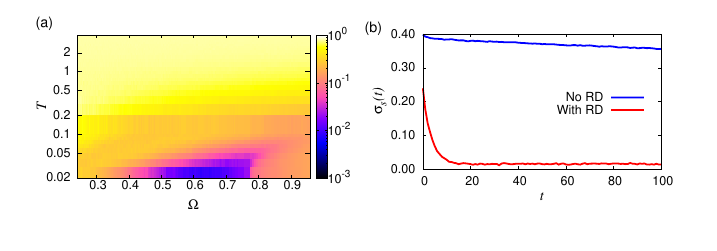}
\caption{
(a) The success probability of the rod diffusion update for an $L=8$
lattice.
(b) Spatial variance of the dimer density (Eq.~\eqref{svd}) as a function
of Monte Carlo time with and without the rod diffusion update for $L=4,T=0.2$
and $\Omega=0.6$.
}
\label{FigRD2}
\end{figure}

To understand the impact of this update intuitively, let us first consider
the limit of large $\Omega$. In this limit, the rod coverage is low, and
we do not expect a dense packing of rods. Rod movements will statistically
encounter no conflicting arrangements, and the update will terminate within
a small number of moves. In the opposite limit of zero $\Omega$, rods do
not have any end points, and the condition of more than one conflicting rod
is never met as every
spatial site is occupied by at most one time-spanning rod, and nearest
neighbors cannot be simultaneously occupied (as in Fig.~\ref{FigRD1}(d)).
The space-time
configuration can thus be collapsed to only a spatial configuration, which
is the natural compression of the path integral in the absence of quantum
fluctuations, and considered purely as a dimer covering. Our update thus
reduces exactly to a worm update for a classical dimer model, and has
a success probability of unity in this limit. This implies that for the
two easily understandable limits, we can expect the update to have a
high success probability, but in the intervening region this probability
may be suppressed due to high incidences of having strictly more than one
conflicting rod in all possible directions of movement.
This would correspond to highly jammed configurations
which cannot be easily updated.

As explained for the previously described updates, we estimate the region
of space-time visited by a single update on average without regarding
whether the update is successful or not, and use this estimate to calculate
the number of repetitions required to fully sample space-time. Now we
can estimate the fraction of space-time which is successfully updated
in one Monte Carlo step by our update, and use this to understand the
efficiency of the update. We show numerical results for the same in
Fig.~\ref{FigRD2}(a)
for a range of temperature $T$ and $\Omega/\delta$, and we
find that indeed in the two limits discussed above and for values
of temperature $\ll\delta,\Omega$,
our update has an $O(1)$ success probability. As shown in Fig.~\ref{FigRD2}(a),
the inefficiencies begin to be visible only at low temperatures and
intermediate values of $\Omega$.

This update is critical for changing spatial patterns in configuration space.
To highlight the importance of this, we can study the decay of a random fixed
pattern chosen at equilibrium, and then study the equilibration process with
and without this update. We do this by considering the spatial variance,
defined as
\be\label{svd}
\sigma^2_s(t)=\frac{1}{N}\sum_i (\braket{n_i}_{av}-\bar{n})^2,
\ee
with $\bar{n}=\frac{1}{N}\sum_i\braket{n_i}_{av}$, where $av$ denotes averaging
over a large number of Markov chains.
This is calculated as a function of Monte Carlo time.
To analyze the efficiency of this update, we choose $\Omega=0.6$ and
$T=0.2$, as this choice of parameters belongs to a regime where
the rod diffusion update has a high relative probability (Fig.~\ref{FigRD2}(a)),
and the other
updates are not efficient (seen in acceptance probabilities in
Figs.~\ref{FigLS}(b),\ref{FigVS}(b)).
Our results are
presented in Fig.~\ref{FigRD2}(b), and we see that the local updates are not
able to lose memory of the initial condition as the variance does not decay
significantly, whereas using the rod diffusion update the variance drops
to zero rapidly. The small residual value of the variance is due to the
finite number of Markov chains which we have used (as $\braket{n_i}_{av}
=\bar{n}$ only in the limit of averaging over an infinite number of Markov
chains).

An extension of the worm update to spatio-temporal regions which do not
contain quantum fluctuations has been implemented for the quantum dimer
model~\cite{dabholkar2022reentrance}. Additionally, a different version of
the worm update (where the update sweeps across all of imaginary time
without any breaks) has recently been utilized for the Rydberg atom
array on the Kagom\'e lattice~\cite{wang2024renormalized}.
These updates are useful at
intermediate temperatures and are closely related to the rod diffusion update
discussed above. Toggling of rods on local motifs (such as pairs of sites)
has also been used for a frustrated transverse field Ising model
~\cite{hearth2022quantum},
and this would again be accessible as a space-local version of our
updates.

\subsection{Optimized parallel tempering}

As the updates described above work efficiently in different regions of
phase space, it is important to engineer a way to mix configurations from
simulations at different parameter values. One of the most commonly used
ways to do this is by parallel tempering \cite{hukushima1996exchange},
where simulations (replicas)
are simultaneously run at slightly separated parameter
values, and space-time configurations are exchanged using the detailed
balance condition
\be
\frac{P(\{A_1,A_2\}\to\{A_2,A_1\})}{P(\{A_2,A_1\}\to\{A_1,A_2\})}=
\frac{P_1(A_2)P_2(A_1)}{P_1(A_1)P_2(A_2)},
\ee
where $P_{1(2)}$ corresponds to the probability distribution
and $A_{1(2)}$ to the current operator string configuration 
at parameter set $1(2)$. It is desirable to have a parameter spacing
such that configurations living in the limits of phase space where we
expect our updates to lead to fully ergodic behavior are able to exchange
their positions, thus leading to a high degree of mixing at all intermediate
parameter values.
However, as system size is increased, the spacing between
neighboring parameter sets needs to be reduced to ensure a high rate of
exchange from the detailed balance condition. This requires an increasing
number of parameter values at which simulations must be run, leading to
limitations on size due to computational resources.

\begin{figure}[t]
\begin{center}
\includegraphics[width=0.6\hsize]{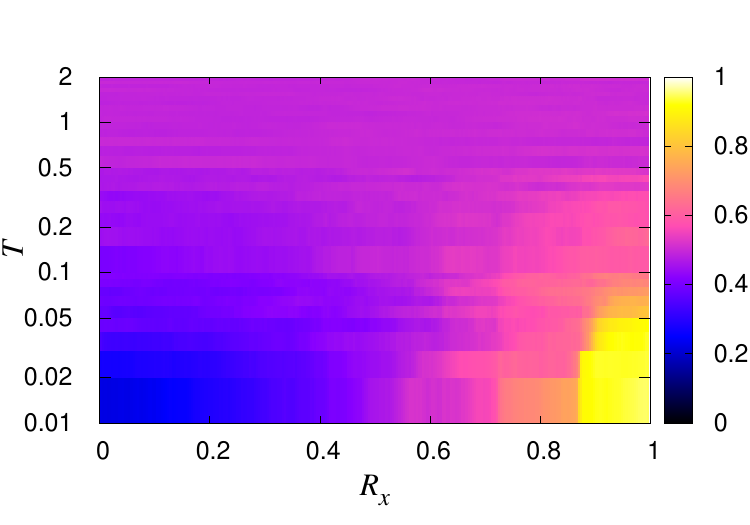}
\end{center}
\caption{
The average replica number at each point in parameter space 
(replica index $R_x$ as defined
in the text). The replica index is normalized to be between 0 and 1.
At high temperature we see that at each location, all replicas appear
over the course of the entire simulation. The data shown above has been
obtained for an $L=4$ system.
}
\label{FigPT}
\end{figure}

It is known from free energy considerations \cite{hukushima1996exchange}
that in regions of parameter space where free energy varies slowly,
it is possible to have a good exchange rate even if parameter spacing
is large. Small spacing is thus only required close to phase transitions.
Following Ref.~\cite{katzgraber2006feedback},
we use a feedback optimized mechanism to choose the parameter spacing
to ensure that we have even diffusivity of the replicas at all points
in phase space. Our optimization procedure is carried out at a fixed
$T$ and varying $\Omega$ for $\delta=1$. We fix an upper and lower
value for $\Omega$ such that we are sure that our updates will provide
good ergodicity in these two limits, and choose 254 intervening values
$\Omega_i$, whose locations can be optimized. This requires a total of 256
parallel simulations, and this number is chosen as the computer hardware at
our disposal allows efficient parallelization over 256 cores for each node.
The success of the optimization procedure can be quantified by measuring
the fraction of replicas which have most recently visited one of the
ends of the parameter range. In practice we assign a label ``up''(``down'')
to replicas which have most recently visited the upper(lower) end. At each
parameter value, we calculate histograms $n_{up}$ and $n_{down}$, which
quantify the number of replicas which lived on this parameter value with
each label. For convenience, one can now express the number of replicas
with the down label as a fraction,
\be
f(\Omega_i)=\frac{n_{down}(\Omega_i)}{n_{up}(\Omega_i)+n_{down}(\Omega_i)}.
\ee
The locations of $\Omega_i$ are chosen to maximize the current of replicas,
and if the optimization procedure is successful, one expects
$f(\Omega_{i+1})-f(\Omega_i)$ to be constant for all $i$. Although this
ideal scenario is not reached in our optimization, we find that the
variations are small, and that at all values of $\Omega_i$, we have
replicas which travel to both ends of the parameter range. This is
sufficient to ensure ergodicity.

For the parameter range we have picked, we exchange nearest neighbor
pairs, i.e. replicas living at $\Omega_i$ and $\Omega_{i+1}$.
At each Monte Carlo step, we alternate between exchanging replicas
amongst odd or even numbered pairs. This implies that at a particular
simulation step, we attempt to exchange $A_1\lra A_2, A_3\lra A_4, ...$,
and on the next step, we attempt $A_2\lra A_3,A_4\lra A_5, ...$.
We can study the efficiency of our optimization by tracking how
far replicas are able to move within a set simulation time.
We denote the initial location of the replica by $R_x$.
This would naively be an integer between $0$ and $255$, and we
normalize it to be in the interval $[0,1]$. We calculate the
average location of each replica over the entire duration of
a simulation, and plot it as a function of
$R_x$. Our results are shown for a range of $T$ in Fig.~\ref{FigPT},
and we see that at high temperatures, the average location of every
replica is close to $0.5$, and only at $T<\Omega_{mean}/10$, the
average location is correlated with the initial position.
The data presented in Fig.~\ref{FigPT} is for $L=4$, and using the
limits for $\Omega$ as 0.24 and 1.00. For these values, we have
already shown in Fig.~\ref{FigRD2} that we have high efficiency
for the rod diffusion update.
Note that the optimized parallel tempering  which we have discussed
above is not efficient at low temperatures, and thus we need updates
which are able to sample efficiently in each regime separately. Thus,
using just a combination of parallel tempering and simple local updates
would be insufficient at low temperatures.

\subsection{Benchmarking and autocorrelation functions}

\begin{figure}[t]
\includegraphics[width=\hsize]{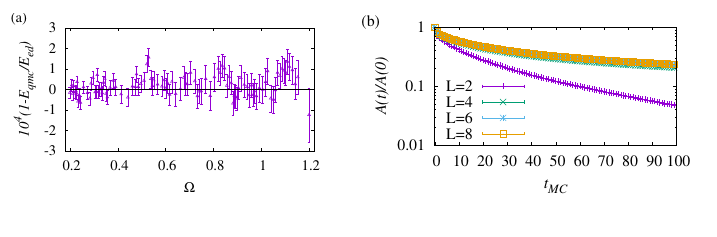}
\caption{
(a) Relative difference between energies using exact diagonalization and
QMC. The data has been obtained using $L=2$ and $T=0.006$, and the difference
is amplified by a factor of $10^4$ to show agreement within error bars.
(b) Autocorrelation function as defined in Eq.~\eqref{eauto}, as a function
of Monte Carlo time, for various system sizes at $T=0.05$.
}
\label{FigBench}
\end{figure}

To ensure that we have implemented the detailed balance conditions correctly,
we perform a comparison with exact diagonalization (ED) for a small system size.
For the $L=2$ periodic lattice shown in Fig.~\ref{FigKL},
the number of dimer occupation states allowed by the Rydberg constraint are
$2649$. This number is small enough to perform a full diagonalization, and
thus to calculate energy at finite temperature. We carry out a comparison
of the energy for a range of $\Omega$ as shown in Fig.~\ref{FigBench}(a),
and find that the agreement between ED and QMC is of the order of $0.01\%$,
and within the error bars of the QMC simulation. Thus we can conclude that
the QMC simulations are unbiased and accurate. Note that in Fig.~\ref{FigBench}(a),
we have used $T=0.006$, which is $\ll\Omega$ for the range of $\Omega$
studied.

As energy is a coarse grained observable, it is also important to study
correlations of the space-time structure to understand how effective our
updates are modifying the same.

We do so by considering auto-correlations
of the dimer occupation, defined as
\be\label{eauto}
A(t)=\sum_{\tau}\frac{1}{N}\sum_i\braket{n_i(\tau+t)n_i(\tau)}-\braket{n}^2,
\ee
where $t,\tau$ denote simulation time, and $n=\frac{\sum_i n_i}{N}$ is the
average dimer density. The autocorrelation is calculated using the
spatial configuration on a random reference slice, and the location of
this reference slice is kept fixed for all simulations.

As we are running a large number of replicas in parallel, and exchanging
them at every Monte Carlo step, we calculate the autocorrelation for
each replica, by just considering its spatial pattern.
Thus we observe how the pattern de-correlates as it moves over parameter
space. As we expect all replicas to visit all of parameter space, we
average over the autocorrelation functions of individual replicas,
and thus improve the associated statistics.
The data for the autocorrelation is only collected
after the simulation has had a sufficient number of warm up steps, and is
sampling from the equilibrium distribution.
As we would like to know the performance of our
algorithm at low temperature and intermediate values of $\Omega$, we
fix $T=0.05$ and $\Omega\in[0.24,1.00]$. The autocorrelation for various system sizes
is shown in Fig.~\ref{FigBench}(b),
and we see that $A(t)/A(0)$ decays rapidly for $L=2$, and saturates for the
3 larger system sizes, namely $L=4,6,8$.
This result suggests that the algorithm is fairly efficient
at sampling even for $T\ll\Omega$ for the most challenging range of $\Omega$
expected.

\section{Phase diagram of Kagom\'e link lattice}

We now study the Kagom\'e link lattice using ED for the
$L=2$ (24 sites) periodic lattice to gain
an understanding of the phase diagram and study larger system sizes
using the QMC updates discussed in the previous section. The latter allow
us to extract the phase diagram in the thermodynamic limit.
The Rydberg atom array on this lattice has been discussed in
Ref.~\cite{Vryd},
where it was found that in the limit of $\Omega\gg\delta$, the ground state
is a trivial paramagnet in the language of spin-1/2 particles,
and that by lowering
$\Omega$, one can cross into a quantum spin liquid phase and eventually into a
valence bond solid phase with a large unit cell. These conclusions were
reached by carrying out DMRG simulations on quasi-1D cylindrical geometries,
and we would like to study the temperature dependence using our QMC directly
in the 2D limit. The effect of temperature is most easily understood in the
classical limit, where at high temperature, the system accesses all Rydberg
allowed configurations, and at $\beta\gg\delta$, the system only accesses
the manifold of states which have maximal density. This set is still
macroscopically degenerate due to the inherent frustration of the underlying
lattice. Once a quantum fluctuation is added, we expect a second energy scale
controlled by both $\Omega$ and $\delta$, where the residual classical entropy
is released, leading to a ground state with at most $O(1)$ degeneracy.
In the phase diagram discussed below, we thus expect to find four phases.
At high temperature, we expect a simple paramagnetic phase with maximal
entropy and no signature in the order parameters we use to distinguish
the other phases. At low temperature and $\Omega\gg\delta$, we expect
the quantum paramagnet, which differs from the the high temperature one
only in terms of entropy as it should have zero entropy for temperatures
below the energy gap. At $\Omega\ll\delta$, we expect the classical spin
liquid for $\Omega<T<\delta$, which has a non-zero entropy which is lesser
than the maximal possible. In addition to the entropy, we also use a string
order parameter to distinguish between the classical spin liquid and the
quantum paramagnet as the former shows a non-zero value for strings of
significant size whereas the latter does not. The quantum spin liquid
is expected to emerge when we lower the temperature to $T\ll\Omega$ starting
from the classical spin liquid regime, and this is characterized by zero
entropy and a string order parameter behavior similar to that of the
classical spin liquid. We do not find evidence of this last phase in
the phase diagram which we are able to generate using the QMC simulations.

\begin{figure}[t]
\includegraphics[width=\hsize]{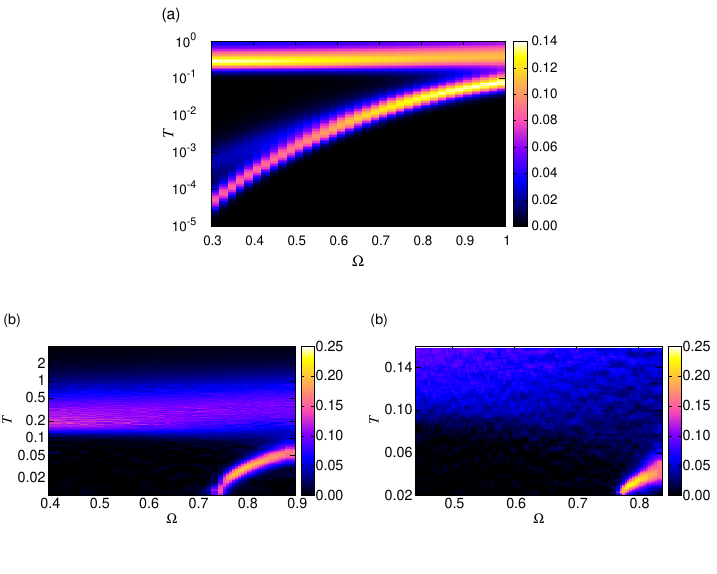}
\caption{Specific heat for the Kagom\'e link lattice :
(a) the $L=2$ case, simulated using exact diagonalization, shows a smooth shift
of the lower peak with $\Omega$;
(b) the $L=4$ case with QMC, where temperatures as low as $0.01$ can be reached,
shows a behavior which is qualitatively consistent with the $L=2$ case.
(c) the $L=8$ case, again similar to the $L=4$ case.
}
\label{FigData}
\end{figure}

Using ED, we study the specific heat as a function of both $\Omega$
and $\beta$ (for convenience we set $\delta=1$). We can trace the lower
release in entropy as a peak in the specific heat, and we show this in
Fig.~\ref{FigData}(a). We find that this peak moves smoothly from $T\approx 10^{-5}$
at $\Omega=0.3$ to temperatures of $O(1)$ at high $\Omega$.
This suggests that the curve traced by the location of the peak in
Fig.~\ref{FigData}(a) ends at $T=0$ for $\Omega=0$, which would be the
natural expectation. At
$\Omega\approx 1$, the distinction between the two peaks in specific heat
which signify the release in entropy due to classical constraints and
quantum fluctuations respectively, is lost.
This is expected for $\Omega\approx\delta$
as the separation between scales required for two distinct peaks is lost.
This phase diagram suggests that the quantum paramagnet is the correct
ground state for all $\Omega>0$. This is in contrast with the expectation
of a quantum spin liquid phase. We find that the low lying spectrum for small
$\Omega$ has a unique ground state with a roughly equally spaced spectrum
above it. This is also inconsistent with a $Z_2$ spin liquid on a torus,
where one would expect four low lying states which have a topological nature,
and are well separated from the rest of the spectrum.

%As the specific heat
%peak at intermediate temperatures suggests a release in entropy due to the
%constraint of fully packed dimers, we can consider the ground state at
%$\Omega\ll\delta$ to be a superposition over fully packed dimer states.
%To identify the similarity with the uniform superposition seen in the
%ground state of the ideal toric code, one can measure the resonances along
%closed strings using a product of $\sigma^x(=b+b^{\dagger})$. We show the
%loop we use for this measurement on a 24-site lattice in Fig.~\ref{FigED},
%and the expectation value $Tr[e^{-\beta H}\Pi_{i\in loop}\sigma^x_i]$ as a
%function of $\Omega/2$ and $T$ in Fig.~\ref{FigED}. We find that for
%$\Omega<0.7$, there is a low temperature phase bounded by the location of
%the specific heat peak in temperature, where this expectation value
%is $O(1)$, which is comparable to value of unity expected for a perfectly
%uniform superposition.
%This suggests that in this parameter regime, the ground state
%indeed shows resonances which are consistent with a $Z_2$ spin liquid.
%Note however that this resonance is stable to temperatures order of magnitude
%smaller than the smallest scale ($\Omega$) in the problem, implying high
%sensitivity to thermal fluctuations.

The QMC simulation setup can now be used to extract the residual entropy
as a function of temperature and $\Omega$. As our simulations are
restricted to sample only from the blockade allowed states, the thermal
entropy at infinite temperature cannot be easily calculated from product
states, as dimer coverings cannot be represented in such a manner.
To calculate a benchmark for the infinite temperature entropy, we
use specific heat data at large $\Omega(\approx 1.4)$, and integrate
$C_v/T$ starting from temperatures much below the expected gap. As the
system is in the paramagnetic phase in this limit, the ground state
residual entropy is trivially expected to be zero, and the integration
to high temperatures allows us to extract the infinite temperature
entropy. We find that the entropy per unit site is
$0.3285(1)$ for $L=4$, and has negligible finite size corrections
for larger sizes. As expected, this is significantly
smaller than $ln(2)$, which would be the case without the Rydberg
constraint.

We show our results in Fig.~\ref{FigData}(b) and (c),
and find that these are similar to the $L=2$
lattice for system sizes of $L=4$ and $L=8$ down to temperatures of
$\approx 0.01$.
For low temperatures, calculating $C_v$ using the fluctuation-
dissipation relationship, i.e. $C_v=\beta^2(\braket{E^2}-\braket{E}^2)$,
has large error bars, and thus we obtain $C_v$ directly by
taking a numerical derivative of the energy. For our simulation
ranges, this ensures an error bar of $<10\%$.
The smallest value of $\Omega$ at which this temperature is low
enough to detect the second peak in specific heat is $0.7$, and the
temperature considered is 30 times smaller than the smallest scale in the
system ($\Omega/2=0.35$). This point is consistent with the location of
the transition to the spin liquid phase in Ref.~\cite{Vryd}.
If the spin liquid phase indeed appears close to this transition, it is
not seen down to temperatures much smaller than $\Omega,\delta$.
%We also study the density of dimers $\langle n\rangle$ as a function of
%$\Omega$ at the lowest temperature simulated, and find a peak in its
%derivative with respect to $\Omega$ at the same
%position seen in Ref.~\cite{Vryd}.
%However, in our case, it is a smooth peak as it is finite temperature
%and does not sharpen into a discontinuity.

\begin{figure}[t]
\includegraphics[width=\hsize]{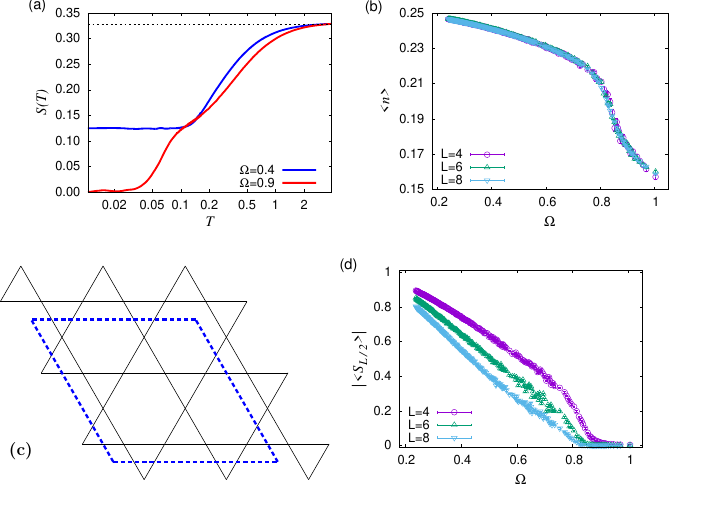}
\caption{
(a) Entropy for the $L=4$ case, calculated using specific heat data of Fig.~\ref{FigData}, for $\Omega=0.2$ and $0.45$. Expected infinite temperature entropy
shown by dashed line.
(b) Dimer density $<n>$ as a function of $\Omega$ for $L=4,6$ and $8$ shows smooth behavior at $T=0.05$.
(c) Contour for calculating string order parameter $S_{L/2}$ for $L=3$.
(d) $S_{L/2}$ as a function of $\Omega$ for $L=4,6$ and $8$ at $T=0.05$.
}
\label{FigData2}
\end{figure}

To show that the low temperature phase at small $\Omega$ which we see
in our simulations is a classical spin liquid,
we use the residual entropy and correlation
functions. The former is shown in Fig.~\ref{FigData2}(a) for
$\Omega = 0.4$ and $0.9$, and calculated using the reference
$S(T\to\infty)=0.3285(1)$ as discussed above. We see that at
low temperatures for $\Omega=0.4$, we have a residual entropy,
which should eventually drop to zero below a second crossover
temperature which is too small to access currently.
We define a dimer correlation function as
\be\label{DCorr}
C(r)=\frac{1}{L^2}\sum_{i_x,i_y}\sum_{d=1}^6(\braket{n^d_{i_x,i_y}n^d_{i_x+r,i_y+r}}
-\braket{n}^2),
\ee
where $d$ indexes the position within a unit cell, and $i_x$ and $i_y$ the position
of the unit cell. For the classical
dimer model on the Kagom\'e lattice, this correlation function is known to be
exactly zero past nearest neighbors~\cite{misguich2002quantum}. As the classical
spin liquid in our model can be thought of as the same with monomers, we expect
that we should have an exponential decay of $C(r)$ and we find that this is
indeed the case in our simulations, where we find $C(r)$ vanishes up to an
error bar of $10^{-3}$ for all $r$.

The crossover between the spin liquid phase and the quantum paramagnet can be
captured using the dimer density ($\braket{n}$) and 
indirect observables such as string order parameters. 
We fix $T=0.05$ for our analysis and first show our results for $\braket{n}$
as a function of $\Omega$ in Fig.~\ref{FigData2}(b).
We find a rapid decline in dimer density as
the ground state crosses from the spin liquid to the paramagnetic regime.
However, this decline is smooth and stays unchanged with increasing
system size (Fig.~\ref{FigData2}(b)). Note that although this is expected
for a crossover, this should sharpen into a non-analytic point for a
transition, as seen in the DMRG data of Ref.~\cite{Vryd}.
To understand the spin liquid behavior better, we now turn to string order
parameters.
As our simulations are carried out in the dimer occupation basis, we have
easy access to the $P$ order parameter of Ref.~\cite{Vryd}, and use the
contour shown in Fig.~\ref{FigData2}(c) (this is the same as Fig.6 of
Ref.~\cite{Vryd}). $S_r$ is simply defined as the product of $f(l)$
along all links ($l$) cut by the contour, where we take $f(l)=-1$
if $l$ is occupied by a dimer and $1$ if not, and the contour has size
$r\times r$. For a fully packed dimer configuration, $|\braket{S_r}|=1$,
where the absolute value is taken to negate the sign oscillation due to
odd or even $r$. However, since we have a non-zero monomer density at
all $\Omega$, $\lim_{r\to\infty}\braket{S_r}\to 0$ necessarily in both phases.
For small $r$, we can still expect non-zero values in the spin liquid phase
and this can act as a marker for the crossover to the quantum paramagnet.
This motivates us to study $\braket{S_r}$ for $r=L/2$ for $L=4,6,8$ systems
at $T=0.05$ and we show our results in Fig.~\ref{FigData2}(d).
We see that these order parameter take $O(1)$ values in the spin liquid
phase and rapidly decay around the crossover. As expected, $\braket{S_r}$
decays with increasing $r$ even in the CSL regime.

These studies do not deny the existence of a spin liquid phase,
and although confirming this would require
arbitrarily low temperatures, our results show that the assumed gap
to this spin liquid is smaller than $10^{-2}$ of the microscopic length scales.
This implies that thermal fluctuations will play an important role when
attempting to realize this spin liquid in a Rydberg atom array.

%Figures should only occupy the strictly necessary space, in any case individually fitting on a single page. Each figure item should be appropriately labeled and accompanied by a descriptive caption. {\bf SciPost does not accept creative or promotional figures or artist's impressions}; on the other hand, technical drawings and scientifically accurate representations are encouraged.

\section{Conclusion}

We have presented a construction of cluster updates for stochastic series
expansion QMC for Rydberg atom arrays using a mapping to a hard rod model.
The mapping involves
vertical rods with an exclusion radius, and is applicable for
Rydberg atom arrays on any lattice.
These updates are designed to work efficiently in different parameter
regimes, and we combine them together using an optimized parallel tempering
mechanism. We have studied the efficiency of various updates individually
and also considered their combined capabilities by studying the
autocorrelation of spatial occupation profiles.
Previous attempts to design cluster updates for Rydberg systems
have relied on a mapping to quantum dimer models
\cite{yan2023emergent,yan2022triangular}, whereas our
updates do not have such a constraint and are expected to be
able to sample in varying parameter regimes.
%The rod diffusion update which we have presented here is especially
%well suited to patterns seen in $Z_2$ spin liquids, as resonances within
%such a state would be expected to manifest as equal time rod movements,
%which, when projected on the spatial plane, look exactly like a worm
%update.

The algorithm is applied to study the Kagom\'e link lattice, where a
gapped $Z_2$ spin liquid is expected at low temperatures. Our analysis of
the specific heat provides an upper bound on the gap. This bound is
estimated to be $<\Omega/50$, for the range of the coefficient of the
quantum fluctuation, $\Omega$, where the spin liquid is expected to exist.
This implies that the phase is highly sensitive to thermal fluctuations.
As our results are preliminary, future studies will involve studying
lower temperatures, and other lattices with similar frustration
properties which may lead to more robust spin liquid ground states.

The rod model can also independently
be analyzed from the perspective of renormalization group to understand
analytically the physics of Rydberg systems. There are numerous examples
of interesting phases being realized in such systems, and particular
manifestations which can directly be analyzed using our approach are
a realization of $Z_N$ symmetry in coupled Rydberg chains\cite{sarkar2023quantum}
and a spin glass realized without disorder\cite{yan2023emergent}.
The sweeping cluster update has been used to simulate a paradigmatic
example of a $Z_2$ spin liquid, namely the toric
code~\cite{wu2023u}, and our 
rod diffusion update can be thought of as carrying out the
same task within the Rydberg constrained space.
This is shown explicitly in the classical limit, where we recover the worm
update which samples between configurations of the classical dimer model,
and which are precisely the low energy excitations which one would expect
to be generated by the quantum fluctuations. This suggests that this
update can also be adapted to $SO(N)$ systems on frustrated lattices,
which are expected \cite{block2020kagome}
to host a stable $Z_2$ spin liquid for large $N$.

\section*{Acknowledgements}
We would like to thank Owen Benton for numerous discussions, and Fabien
Alet and Kedar Damle for pointing
out relevant references. Computational resources for this project were
provided by MPIPKS.
The formatting of the pseudo-code was done following Ref.~\cite{seymour2024algorithm}.

\begin{appendix}

\section{Details of local segment update}\label{AppA}

Following the brief description of this update in Sec.~\ref{SLS} and
the representative schematic shown in Fig.~\ref{FigLS}(a), we present
the details and a pseudo-code for the same in this appendix.

First we have a micro-canonical update, which switches between two
space-time configurations which have the same probability. This proceeds by
first picking a random spatial site and a random slice and searching
sequentially in slices above and below the chosen one, until we encounter 
slices in both directions which have either an operator
on the site of interest. In the time direction,
if we isolate the spatial site of interest 
and look at only the operators living on this site, then these operators are
next to each other in the imaginary time direction.	
If either of them is an $H_V$ operator, the update is aborted. For
comparable values of $\delta$ and $\Omega$, we would expect that the update
would reach this stage with a probability of $O(1)$. Once this pair of
neighboring operators is identified, we identify if the state of the site
for the time region between these two end points hosts a rod or not. If it
does, then we can flip this segment with unit probability as we exchange
only between $H_K$ and identity, and both have the same coefficient in the
Hamiltonian. However, if the segment does not host a rod, we must check the
status of the neighboring sites to ensure that in the time region of interest,
they are not occupied by a rod. If they are, then flipping this segment will
create two neighboring rods, which is not allowed by our basis states, and
we must abort the move. After
this check, if we find that we can occupy the segment with a rod without
creating violations of the Rydberg constraint, then we do so with unit
probability. These processes are shown with an example in Fig.~\ref{FigLS}(a),
for rods labeled 1-3. Identity ($I$) operators are represented using
circles on a single site, as these operators do not have any dependence
on the state. Pseudo-code for this process is presented below.

\begin{lstlisting}
__global__ function Local_Segment(int *opr_string)

    int j,k;
    int s_i = random.site();
    int t_i = random.slice();
        k = t_i +1; while(site of opr at k not s_i) {k++;} t_top=k;
        k = t_i -1; while(site of opr at k not s_i) {k--;} t_bottom=k;
    if (opr at t_top = H_V) or (opr at t_bottom = H_V) : abort;
    else :
        k = rod.starting.upwards.from(t_bottom);
        if (k = no rod) :
        (1) toggle operators (H_K <-> I)at t_top and t_bottom;
        (2) update rod.starting.upwards.from(t_bottom);
        else :
            j = check.if.rod.on.neighbors(s_i,t_bottom,t_top);
            if (j = yes) : abort;
            else : repeat (1) and (2);
            end if
        end if
    end if

\end{lstlisting}

%Pick random spatial site $s_i$ and slice number $t_i$
%Identify first slice with operator at $s_i$ above $t_i$, label $t_{top}$
%Do the same below $t_i$, label $t_{bottom}$
%If operator at $t_{top}$ or $t_{bottom}$ is of type $H_V$ : abort
%Else :

The second process which we use within this update is a canonical move which
toggles between configurations with different probabilities. We again identify
operators at a randomly chosen location on either side of a chosen slice,
and proceed with the update only if both operators are of either type $H_V$
or of type $H_K$. If so, then we identify whether or not the segment
intervening is covered by a rod segment or not. The presence of a rod segment
implies that we can flip the status of a segment without checking its
neighbors. However, as we are allowing ourselves to toggle only between
$H_K$ and $H_V$ as endpoints, if the endpoint is of the form $H_K$, the
resulting arrangement (no rod coverage on both sides of an endpoint) is
inconsistent with an $H_V$ operator. Thus if we have a rod covering the segment
of interest, we require that the endpoints must host $H_V$ operators. If this
condition is not met then we abort our update. If not, then we flip the
status of the segment and the endpoints with probability
$\delta^2/(\Omega/2)^2$. Now turning to the case where the segment does not
host a rod segment; the end points now must necessarily be of the $H_K$ type,
as $H_V$ operators require rod coverage on both sides. Before flipping
we must check if the flip will create inconsistencies with neighboring rods,
if not, then we can accept this flip with probability $(\Omega/2)^2/\delta^2$.
The operations described here are illustrated with a simple example in
Fig.~\ref{FigLS}(a) with rods labeled 4 and 5.
The pseudo-code for this process is almost identical to the one presented above
except for the toggling of $H_K\leftrightarrow I$, which now has an acceptance
probability associated with it (similar to the standard metropolis update for
the classical Ising model).

\section{Details of vertical shuffle update}\label{AppB}

Here we discuss the implementation of the vertical shuffle update (Sec.~\ref{SVS})
along with a short pseudo-code for the same.

First we identify a random site, and run through the entire time direction
to identify the regions where we can move rod segments and segment breaks
freely. We call this as the ``active'' region and an example is highlighted
with a blue border in
Fig.~\ref{FigVS}(a). The entire active region on the site in interest is
the union of all regions like the one shown, which are on the same site
(this is generated by scanning vertically).
In this process, we also create a list of all the segment
breaks and a larger list of all neighboring vertical pairs in the vertical
direction, both within the active region. Care must be taken in this
procedure to also consider the pair which wraps over the
periodic boundary condition in the time direction.
In addition we also track
the number of neighboring $H_V$ pairs, as we will swap these with $H_K$
pairs in this update. Note that a segment break is defined
in our context to mean two $H_K$ operators on the spatial site of interest,
which are neighbors in the vertical direction and the space between them
is not covered by a rod. In addition, none of the time slices contained between
these two $H_K$ operators should host a rod on any spatial neighbor of
the site under consideration.
If both the numbers of $H_K$ pairs (denoted by $k$)
and $H_V$ pairs (denoted by $v$)
are non-zero, we proceed by picking a random $H_K$ pair and attempting to
exchange it with a random $H_V$ pair. Following the convention developed
in previous subsections, $H_V$ and $H_K$ are denoted by dashed and solid
horizontal lines respectively, and in Fig.~\ref{FigVS}(a) we see that there
are many more of neighboring $H_V$ pairs than $H_K$ pairs.
Although both configurations have the same
probability, one must be careful while carrying out this exchange as the
forward, $A\to B$, and reverse $B\to A$ processes may not have the same
proposal probability (here $A$ and $B$ are the configuration before and
after the update). In other words, to satisfy the detailed balance condition,
$P(A)P(A\to B)=P(B)P(B\to A)$, we must now require $P(A\to B)=P(B\to A)$,
as $P(A)=P(B)$. If configuration $A$ has $k_A$ number of $H_K$ pairs
and $v_A$ number of $H_V$ pairs then $P(A\to B)$ is given by $(k_Av_A)^{-1}$,
as $B$ is defined by picking one out of $k_Av_A$ options.
The simplest way to consistently follow detailed balance now is to require
$k_A=k_B$ and $v_A=v_B$. This is engineered as follows.

We pick an $H_K$ pair randomly and check the operators immediately above
and below the pair at the same spatial location. If both of these operators
are of the $H_V$ type, then we proceed to pick an $H_V$ pair which also
has this property. Now that we have identified two pairs which have the same
``local'' environment, we can proceed with the exchange with unit probability.
The particular check which we have used is one of the easiest ways to ensure
that the condition of $k_A=k_B$ and $v_A=v_B$ described above is satisfied
as the number of $H_K$ and $H_V$ pairs are individually explicitly conserved.
An example of all such $H_K$ pairs is shown in Fig.~\ref{FigVS}(a) in green
coloring.

Since this update scans the entire time span of an individual site, simply
repeating this update $N$ times covers all of space time
on average. For an individual
update, the number of pair exchanges attempted is taken to be the number
of $H_K$ pairs identified, as this allows each pair to move once on average.
The success probability of our update can be quantified by recording the total
length of the pairs which are successfully exchanged
as a fraction of the time span of the system for a single update.
Note that every proposed exchange is not accepted, as the local environment
may not satisfy the condition discussed above. A pseudo-code for this
update is presented below.

\begin{lstlisting}
__global__ function Vertical_Shuffle(int *opr_string)

int j,k,t_j,t_{j+1};
int s_i = random.site();
int opr.list = filter(opr_string,s_i);
int nbr.opr.list[number.of.nbrs.of.s_i];
for j in number.of.nbrs.of.s_i :
    nbr.opr.list[nbr.number(j,s_i)] = filter(opr_string,nbr.number(j,s_i));
end for
int Hk.list, Hv.list;
for j in size.of(opr.list) :
    if (opr.list[j].type = Hk) and (opr.list[j+1].type = Hk) :
        t_j=opr.list[j].t-position;
        t_{j+1}=opr.list[j+1].t-position;
        k = check.if.rod.on.neighbors(s_i,nbr.opr.list,t_j,t_{j+1});
        if k = 0 : #no conflicting rods
            add(j,Hk.list);
        end if
    else if (opr.list[j].type = Hv) and (opr.list[j+1].type = Hv) :
        add(j,Hv.list);
    end if
end for

int pair_k,pair_v;
int opr_k_above,opr_k_below,opr_v_above,opr_v_below;
for j in size.of(Hk.list) :
    pair_k = random_element(Hk.list); #pair indexed by lower opr location
    opr_k_above = opr.list(pair_k+2); #gets opr above pair
    opr_k_below = opr.list(pair_k-1); #gets opr below pair
    pair_v = random_element(Hv.list);
    opr_v_above = opr.list(pair_v+2); #gets opr above pair
    opr_v_below = opr.list(pair_v-1); #gets opr below pair
    if (opr_k_above.type=Hv) and (opr_k_below.type=Hv) :
        if (opr_v_above.type=Hv) and (opr_v_below.type=Hv) :
            exchange(pair_k,pair_v);
            update(opr.list);
            update(Hk.list);
            update(Hv.list);
        end if
    end if
end for

\end{lstlisting}

\section{Details of rod diffusion update}\label{AppC}

In this appendix, we detail the procedure for implementing the rod diffusion
update presented in Sec.~\ref{SRD} along with a pseudo-code for the same.

\begin{figure}[t]
\begin{center}
\includegraphics[width=0.7\hsize]{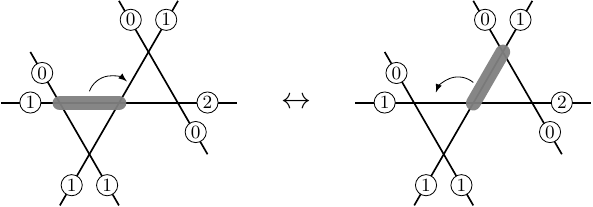}
\end{center}
\caption{
A single move within the rod diffusion update shown by projecting it
onto the spatial plane. Circled are the numbers of rods at each spatial
site which have a temporal overlap with the rod being moved (shown as
gray dimer). This move can proceed as proposed as the number of legal
forward options ($=1$), is the same as the number of legal backward
options ($=1$).
}
\label{FigRDp}
\end{figure}

Each spatial site on the Kagom\'e link lattice can be visualized as a dimer
between two neighboring sites on the Kagom\'e lattice, and it can be moved
by pivoting it about one of its ends, as shown in Fig.~\ref{FigRD1}(a).
By extending
this into the temporal direction, this pivot corresponds to a movement
of a rod. The spatial site of the Kagom\'e lattice about which this pivot
is carried out is chosen to not be in the membrane of modification already
developed, as this would lead to the update undoing its own moves. This is
similar to the spatial worm update \cite{alet2006classical},
where dimers are pivoted around
an exit site, which is not the site via which the worm entered the dimer.
For the initial rod, this site is chosen at random, as there is no pre-existing
membrane. An example of the membrane of modification is shown in
Fig.~\ref{FigRD1}(b).
For the Kagom\'e lattice, a pivoting dimer has three possible options for its
new position, as shown in Fig.~\ref{FigRDp}. To decide which of these options to
choose, we consider the corresponding rod configuration. We calculate the
number of conflicting rods generated by moving the current rod to each of
these three spatial positions while keeping the temporal ends of the rod
the same. We choose the position with the least number of conflicting rods,
and denote this number by $n_c$.
If there are multiple such positions, we store the number of them by denoting
them as the number of forward options ($p_f$), and choose one at random
with uniform probability. If $n_c>1$, detailed balance would require a
reverse proposed move which coordinates the movement of $n_c$ rods into the
relevant positions, leading to the reverse movement of the current rod. This
would imply that we would have to engineer an update which branches, and
carries movements of multiple rods at the same step of the update. This
would significantly complicate the update, and we cannot visualize a simple
way to do this currently. An example of this sort of movement is shown in
Fig.~\ref{FigRD1}(c) and (d), where the movement of the current rod (shown
in red) to its new position (shown in blue) creates a conflict with two
rods. As shown, these two rods can be on the same spatial site or on
different ones.

Thus if $n_c>1$, we abort the update. If not, we consider the reverse move,
where the rod is being moved from its future position to its current
position. In the dimer language, this is now the reverse pivot. Following
the same procedure to pick the new position of the rod leads to a
calculation of the number of backward options $p_b$. Detailed balance now
requires that the probability of the forward process, $1/p_f$,
be equal to that of the reverse process, $1/p_b$. This trivially translates
to requiring $p_f=p_b$. If this condition is not met, we abort our update.
We relax this condition only for the initial and final rod movement, where
we require that $p_f^{start}=p_b^{end}$ and $p_b^{start}=p_f^{end}$.
A sample pivot process which satisfies detailed balance is shown in
Fig.~\ref{FigRDp}. If the update is accepted, we update the relevant
operators. With each rod movement, we carry all operators ($H_K$ and $H_V$
type) which are living on it to the next position inhabited by the rod.
The update terminates when the current rod is in a position where there
are no more conflicting rods to move. A short pseudo-code for this update
is presented below.

\begin{lstlisting}
__global__ function Rod_Diffusion(int *opr_string)

#Prepare rod configuration from opr_string beforehand
#Copy opr_string to temp_opr_string

int j,k,l;
int rod_curr = random.rod();
int pivot = random-end(rod_curr.site); #pivot chosen as one of dimer ends
int t_end = top-end(rod_curr);
int b_end = bottom-end(rod_curr);

#recursion begins, track step number as r-step
int r-step=1;
while(1) : #infinite loop broken when valid config met
    for j in nbrs(rod_curr,pivot) : #positions to which the dimer can pivot
        number-of-valid-exits[j]=0; #positions with conflicts<2 added here

        # Below, choose pivot point for j which is other end of dimer j
        temp-pivot=dimer-end(j,1);
        if (temp-pivot=pivot) : temp-pivot=dimer-end(j,2);

        l=0;
        for k in nbrs(j,temp-pivot) : #counting conflicts if rod_curr -> j
            l+=number-of-rods(k,b_end,t_end); #in range b_end to t_end
            possible_exit[j]=rod-number(k,b_end,t_end);
        end for
        if (l<2) :
            number-of-valid-exits[j]++;
        end if
    end for

    p_f=0; #number of possible forward options
    min-conflict=2;
    for j in nbrs(rod_curr,pivot) :
        if (number-of-valid-exits[j]<min-conflict) :
            min-conflict=number-of-valid-exits[j];
        end if
    end for
    #Above we have preferred the exits with zero conflict over one conflict

    for j in nbrs(rod_curr,pivot) :
        if (number-of-valid-exits[j]=min-conflict) : p_f++;
    end for
    if (p_f=0) : abort;

    int choice_forward = random(p_f); #choose out of p_f options
    
    #Using choice_forward and pivot, now calculate the number
    #of options p_b for the reverse pivot, using same steps
    #as for the forward move

    if (recursion-step=1) :
        p_f_strart=p_f; p_b_start=p_b; #store number of options for start
    end if
    if (number-of-conflicts[choice_forward]=0) :
        p_f_end=p_f; p_b_end=p_b; #same for end
    end if

    if ((p_f=p_b)||(r-step=1)||(number-of-conflicts[choice_forward]=0)) :
        rod_curr=choice_forward;
        update(temp_opr_string,rod_curr,choice_forward); #move operators along with rod
    else :
        abort;
    end if

    if (number-of-conflicts[choice_forward]=1) :
        rod_curr=possible-exit[choice_forward];
        t_end=top-end(rod_curr);
        b_end=bottom-end(rod_curr);
        pivot=dimer_end(rod_curr,1);
        if (pivot in choice_forward): pivot=dimer_end(rod_curr,2);
        #We want the pivot to be away from choice_forward
        #Below, checking detailed balance for start-end
    else if ((p_f_start=p_b_end) and (p_f_end=p_b_start)): 
        exit; #update completes as no conflicting rods
    else :
        abort;
    end if
    r-step++;
end while

if (valid exit without abort) : opr_string=temp_opr_string;

\end{lstlisting}

\end{appendix}

% TODO:
% Provide your bibliography here. You have two options:

% FIRST OPTION - write your entries here directly, following the example below, including Author(s), Title, Journal Ref. with year in parentheses at the end, followed by the DOI number.
%\begin{thebibliography}{99}
%\bibitem{1931_Bethe_ZP_71} H. A. Bethe, {\it Zur Theorie der Metalle. i. Eigenwerte und Eigenfunktionen der linearen Atomkette}, Zeit. f{\"u}r Phys. {\bf 71}, 205 (1931), \doi{10.1007\%2FBF01341708}.
%\bibitem{arXiv:1108.2700} P. Ginsparg, {\it It was twenty years ago today... }, \url{http://arxiv.org/abs/1108.2700}.
%\end{thebibliography}

% \bibliographystyle{SciPost_bibstyle} % Include this style file here only if you are not using our template
\bibliography{scpreferences.bib}

\nolinenumbers

\end{document}